\documentclass{article}
\usepackage{spconf,amsmath,graphicx}
\usepackage{setspace}

\title{Advancing Multi-Accented LSTM-CTC Speech Recognition using a Domain Specific Student-Teacher Learning Paradigm}
\name{Shahram Ghorbani, Ahmet E. Bulut, John H.L. Hansen \thanks{This project was funded by AFRL under contract FA8750-15-1-0205
and partially by the University of Texas at Dallas from the Distinguished University Chair in Telecommunications Engineering held by J. H. L. Hansen.}}
\address{Center for Robust Speech Systems (CRSS) \\ 
University of Texas at Dallas, Richardson, TX 75080\\
\{shahram.ghorbani, ahmet.bulut, john.hansen\}@utdallas.edu}

\begin{document}
%
\maketitle
\begin{abstract}
`Non-native speech causes automatic speech recognition systems to degrade in performance. Past strategies to address this challenge have considered model adaptation, accent classification with a model selection, alternate pronunciation lexicon, etc.  In this study, we consider a recurrent neural network (RNN) with connectionist temporal classification (CTC) cost function trained on multi-accent English data including US (Native), Indian and Hispanic accents. We exploit dark knowledge from a model trained with the multi-accent data to train student models under the guidance of both a teacher model and CTC cost of target transcription.  We show that transferring knowledge from a single RNN-CTC trained model toward a student model, yields better performance than the stand-alone teacher model. Since the outputs of different trained CTC models are not necessarily aligned, it is not possible to simply use an ensemble of CTC teacher models. To address this problem, we train accent specific models under the guidance of a single multi-accent teacher, which results in having multiple aligned and trained CTC models. Furthermore, we train a student model under the supervision of the accent-specific teachers, resulting in an even further complementary model, which achieves +20.1\% relative Character Error Rate (CER) reduction compared to the baseline trained without any teacher. Having this effective multi-accent model, we can achieve further improvement for each accent by adapting the model to each accent. Using the accent specific model's outputs to regularize the adapting process (i.e., a knowledge distillation version of Kullback-Leibler (KL) divergence) results in even superior performance compared to the conventional approach using general teacher models.  
\end{abstract}
\section{Introduction}
\label{sec:intro}

Current successful ASR systems employ deep neural network (DNN) models as acoustic model combined with a hidden Markov model (HMM) \cite{graves2013_hybrid,sak2014}, or use them within an end-to-end configuration \cite{graves2013,deepspeech2}.  Such systems perform well if they are evaluated in the same condition with their training data. However, in real scenarios, speech typically exhibits wide variation due to the differences in room acoustics and reverberation, speakers and accents, and also environment or recording/channel distortions. For many of these situations, it is possible to simulate (augment) or collect more data to generalize the final ASR system. Given training data of multiple conditions, we need to exploit the data efficiently to train an improved multi-condition (domain) ASR model \cite{elfeky2016towards,mirsamadi2017multi, sainath2017no,giri2015improving,yousefi2019probabilistic}. In our scenario, we have data from different accents of English (native US, Hispanic English accent, and Indian English accent) with all other recording conditions remaining constant, thus, allowing us to focus only on the acoustic differences between these accents.

  The general problem of accent within speech technology is a challenging problem, since non-native speech causes loss in performance for speech recognition and diarization systems. The specific problem of accent recognition/classification has been investigated extensively in the past \cite{hansen2016unsupervised,myself,huang2007dialect}. In addition, the notion of accent classification combined to improve speech recognition is also a long-standing goal in the field. Recent advancements in machine learning has caused a renewal in exploring improved techniques to address this problem. There have been many attempts to train a multi-accent (dialect) system. Kanishka et al., in \cite{accent2017} used a multi-task hierarchical CTC-based model with accent-dependent phoneme recognition as a secondary task. \cite{chineseaccent2016} explores an accent-related bottleneck feature as auxiliary information.  \cite{li2017multi} adds a special accent-specific symbol at the end of target transcriptions to train a multi-dialect sequence-to-sequence model. In this study, we investigate employing student-teacher learning to advance a muli-accent model. In student-teacher learning, rather than training a model directly on hard targets, training is carried out in two steps \cite{fukuda2017efficient, hinton2015distilling}. First, we train several complex, distinct,  complementary teacher acoustic models. Next, we train a student model constraining it to mimic the soft-outputs (posteriors or logits) of the original trained teachers. This technique has been successfully applied in the ASR domain: distant-talking ASR \cite{watanabe2017student},  multilingual \cite{cui2017knowledge,hsiao2014improving}, domain adaptation \cite{adaptation}, and others.

In this study, we investigate a novel approach of knowledge transference to achieve an advanced multi-accent model. In our proposed scenario, teacher models are accent-specific RNN-CTC models which are only trained with the corresponding accent English data. However, since the outputs of different trained CTC models are not necessarily aligned, we cannot simply use the ensemble of CTC teacher models. To align the outputs of the accent-specific model, we propose to train them under the guidance of a single multi-accent model. In addition to aligning the CTC trained models, this approach also achieves better generalization. Having the aligned accent-specific models, we train an improved multi-accent model under the supervision of both the teacher models and CTC cost of the target transcription. The resulting multi-accent model significantly outperforms the baseline multi-accent model trained with no teacher model.  Adapting the best multi-accent model to each accent, but constraining it to mimic the soft-outputs of the corresponding accent-specific model generalizes the adaptation process. The proposed approach of adaptation outperforms the knowledge distillation version of KL-divergence \cite{adaptation}.

\section{RNN-CTC Models}

Long Short-Term Memory (LSTM) networks have proven to be effective in many sequential tasks \cite{goodfellow2016deep}. Having gated memory cells to store information within the network enables them to exploit long-range context and produces related outputs with an arbitrary delay. In the speech recognition domain, an LSTM neural network can outperform conventional RNNs \cite{Yu2018}. Alex et al., \cite{BLSTM} introduced a new architecture for LSTM networks that receive input features in both forward and backward directions and showed improved performance for acoustic models compared to a unidirectional trained model. In this study, we also employ bidirectional LSTM cells in the recurrent layers to address the problem of accent/non-native speech recognition. 

Given an input sequence of feature vectors $\textbf{X}=\{x_1, x_2,...\\, x_N\}$, a naive RNN produces a sequence of distributions over a set of output characters $\textbf{Y}=\{y_1,y_2,..., y_N\}$ by iterating the following By intergrating on: 
\begin{equation}
    \begin{aligned}
    \textbf{h}_t = \mathcal{H}(\textbf{W}_{ih}x_t + \textbf{W}_{hh}\textbf{h}_{t-1}+\textbf{b}_h),\\
    \textbf{y}_t = \textbf{W}_{ho}\textbf{h}_{t}+\textbf{b}_o,
    \end{aligned}
\end{equation}

\noindent
where \textbf{W} and \textbf{b} are the weights and biases of the network, respectively, $\mathcal{H}$ is the hidden layer activation function and $y_t$ is the t-$th$ output of the network corresponding to the t-$th$ input ($x_t$). In our scenario, the last layer of our model has  $| \textbf{S}|$ outputs, where  $\textbf{S}=\{characters\; of\;the\;language , ‘blank’, space\\, noise\}$. Here, 'blank' is a special character that is used by CTC for calculating the cost of the output and also by the decoder to output the final sequence.  For each frame, outputs of the model (logits) are submitted to a softmax function to transform them to a valid probability distribution over the members of $\textbf{S}$:
\begin{equation}
Pr(k,t|\textbf{X}) = \frac{exp(y_k^t)}{\sum_{j=1}^{|\textbf{S}|} exp(y_j^t)},   
\end{equation} 
\noindent
where $y_k^t$ is the probability of emitting the \textit{k}-$th$ member of $S$ for the given input $x_t$. We consider the result of the softmax layer for a given sequence $\textbf{X}$ as the matrix $\textbf{O}$ of size $|\textbf{S}|*N$. By choosing one element of each column, we obtain a length $N$ output sequence where its  probability is $ Pr(a|\textbf{X}) = \prod_{t=1}^{N} \textbf{O}(a(t),t)$. The CTC objective is to maximize the probabilities of such sequences that correspond to the target labels:

\begin{equation}
\theta = argmax\sum_{a\in \mathcal{A}} \prod_{t=1}^{N} \textbf{O}(a(t),t).
\end{equation}
Here, $\mathcal{A}$ is the set of all alignments related to the target sequence and $\theta$ represents the parameters of the neural network. Next, given a new input sequence for the trained network, the decoder finds the most probable output character sequence. The study in \cite{graves2013} exploited two decoders: 1) Simply choosing the most probable output from each column of $\textbf{O}$ (best-path); 2) Beam search decoding approach which considers a beam size of N best paths. In our experiments, we employ the second beam search decoding method.

\section{Teacher-Student models for end-to-end CTC speech recognition models}

The first effort to investigate teacher-student model was \cite{ba2014deep}, examining the consequences of having a deep neural network. Hinton et al., \cite{hinton2015distilling}, introduced the term "knowledge distillation", suggesting a new temperature parameter to soften the softmax outputs before being used to guide the training of another neural network. The main idea for student-teacher learning comes from the fact that the distribution of outputs produced by a trained neural network contain underlying relations between output labels. Training another network (student) to output such soft labels which are easier to achieve than hard labels, regularizes the student trained model. As a general setting, having a trained complex model or ensemble of neural networks (teacher models), it is possible to achieve an improved single (smaller) student by constraining it to mimic the soft outputs produced by the teacher(s)  \cite{fukuda2017efficient, hinton2015distilling}. A general framework for teacher-student learning is obtained through the cross-entropy (CE) between outputs of the teacher and student which is represented as: 

\begin{equation}
\mathcal{F}_{CE} = -\sum_{t=1}^{N} \sum_{j=1}^{|S|}\textbf{O}'^{Ref}(j,t) \log(\textbf{O}'(j,t)),
\end{equation}

\noindent  where $\textbf{O}'(j,t)$  and  $\textbf{O}'^{Ref}(j,t)$ are the tempered softmax probability of the $j$-th character at time t for student and teacher model, respectively, which are computed as follows:
\begin{equation}
O'(k,t|\textbf{X}) = \frac{exp(y_k^t/T)}{\sum_{j=1}^{|S|} exp(y_j^t/T)},   
\end{equation}
where T is a temperature. As T becomes larger, the resulting distribution gets softer.

Transferring knowledge from a model trained with CTC is challenging \cite{sak2015acoustic} and has not been adequately explored.  Outputs of a CTC trained model are spiky \cite{spiky}, implying that the model tends to give very sharp posterior probabilities. While the probability of a single class may be close to 1, the rest of the classes are typically closer to 0. In addition, for our setting which uses 'blank', since the model just needs one spike of each character to output the desired transcription, most output spikes are 'blank' which does not have an explicit phonetic similarity with other speech activities, however, there should be some other underlying relation with the neighboring characters.  Finally, because of the fact that CTC does an arbitrary alignment between labels and network outputs, as well as having a model with memory which remembers the acoustic states and outputs spikes at any time, the timing of the probability spikes is different from the true frame-character based alignments \cite{sak2}.

In \cite{sak2015acoustic}, Hasim et al., considered employing student-teacher learning to improve ASR performance for noisy speech with a CTC-based trained model. However, their student model did not outperform the baseline model which was simply trained with noisy data. In that work, they simply used the soft-outputs to train the student model. However, in scenarios where one has the correct transcription, exploiting the target labels in combination with the soft labels would provide greater benefit to the student models \cite{hinton2015distilling}. Therefore, in our student-teacher setting, we use a weighted average of the CE cost of a teacher model and CTC of the true labels as the cost function:
\begin{equation}
L = \lambda  \mathcal{F}_{CE} + (1-\lambda) \mathcal{F}_{CTC}(\textbf{O},\textbf{Y}),   
\end{equation}

\noindent where $\lambda$ is the interpolation weight.

\begin{figure*}[htb]
\centering
  \includegraphics[width=13.7cm]{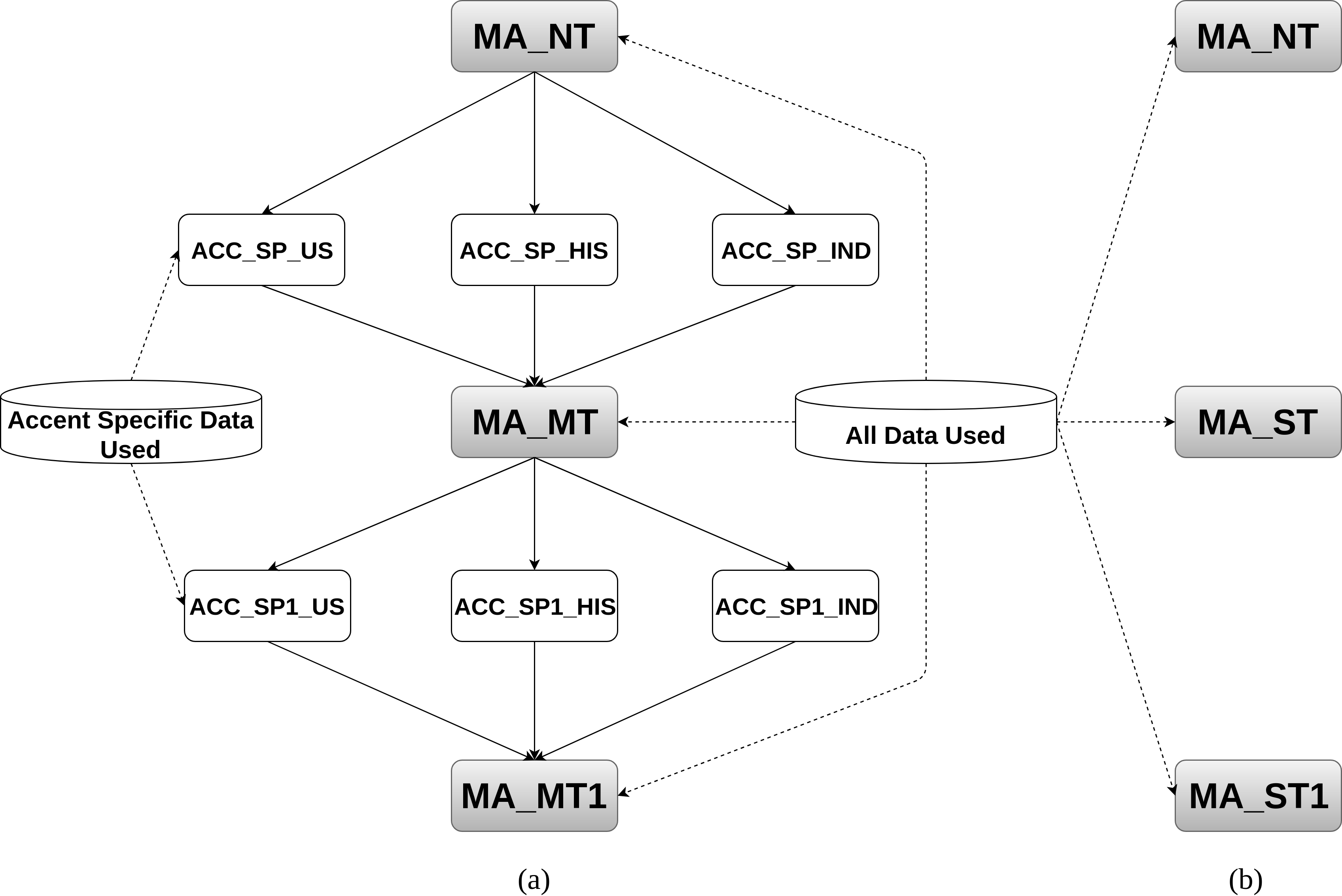}
  \caption{The proposed student-teacher learning to advance a multi-accent model: using accent-specific teachers (a), using multi-accent teachers (b).  }
  \label{fig:framework}
\end{figure*}

Student-teacher learning would be more efficient if it could distill the knowledge from an ensemble of trained complementary models into a single system. It is suggested to construct these complementary models with: training models with different training data, employing alternate architectures of neural network (e.g., convolutional neural network or LSTM), or initializing with different approaches among others. In our case, we consider multi-accent data with the aim to achieve an advanced multi-accent (MA) English model. Each accent has its own underlying similarities between speech acoustic units.  For example, in an Indian accent, phoneme /t/, in many words, is pronounced more like a voiced sound making it closer to /d/. However, in native English, /t/ is an unvoiced phoneme that has the same vocal tract configuration but alternate excitation to phoneme /d/. In addition, for end-to-end ASR that also models the grammar of languages/accents, alternate accents might posses alternate grammar structure or word choice that could influence the relations between acoustic units. 

We hypothesize, accent specific teacher models, which are just trained with data from one accent, benefits the student model more than a multi-accent teacher model.  However, as is discussed, training multiple LSTM-CTC using different accent speech data results in models with their outputs not aligned. To address this issue, we suggest a novel architecture employing student-teacher model: first we train a multi-accent model with training data from all accents, then we train multiple accent specific models from scratch under the guidance of the multi-accent model. Being trained with the same teacher but seeing only the data of one accent results in aligned accent specific models. The last step is to train a new multi-accent model with these individual teachers from which we obtain soft-outputs of each accent from the corresponding model (Figure \ref{fig:framework}).

Having student models that perform better than the teacher provides a new space to explore system advancements. Could these advanced student models teach another generation of students with the aim to achieve further improvement? The knowledge that comes from an improved model is more accurate and probably more close to actual similarities between labels, providing an easier point for students to achieve and generalize the learning. To this end, we apply our proposed student teacher learning one step further as shown in Figure \ref{fig:framework}. We employ the advanced multi-accent model from the previous step (MA\_MT) to train three new accent-specific models (i.e., ACC\_SP1\_US, ACC\_SP1\_IND, and ACC\_SP1\_HIS). Next, we exploit these accent-specific models to teach a further superior multi-accent model (MA\_MT1).

\section{Experimental Setup}

For training and evaluating the model across accents, the UT-CRSS-4EnglishAccent corpus is used \cite{myself}. This corpus of 420 speakers was collected at CRSS-UTDallas and consists of four major English accents: US (native), Hispanic, Indian and Australian. The data for each accent consists of about 100 speakers balanced for gender and age, with session content that consists of read and spontaneous speech. In our study, we use US, Hispanic and Indian parts of the corpus to train both multi-accent and accent-specific models. In this corpus, for each accent, there is about 28h of training data, 5h of development and 5h of evaluation data. 

We extract 26 dim Mel filterbank coefficients for each 25ms frame with a skip rate of 10ms. We expand each frame by stacking 4 frames to each side, then the frames are decimated by skipping 2 frames per each frame for processing. we use the skip process as described in \cite{sak2015fast}.

The neural network architecture starts with two feed forward layers each of 500 neurons, where their outputs go through two bidirectional LSTM layers with 300 neurons in each direction. The LSTM layers are followed by two forward layers each containing 500 neurons. We use Adam Optimizer with an initial learning rate of 0.001 to train the model. Gradients are computed from Mini-batches of 30 utterances. We employ early-stopping by monitoring the performance on a held-out validation set during training epochs. In the evaluation step,  we employ a beam search decoding \cite{graves2013} with a beam width of 100 with no language model or lexicon information. 

\section{Results and Discussion}

\subsection{Knowledge transferring for CTC-based models}
\label{ssec:knowledge_trans_CTC}
In this section, we examine our proposed approach to address the problem of aligning the CTC trained models. The baseline model for this section is trained with the multi-accent data (i.e., US, Indian and Hispanic English) which is used as a teacher model. Next, we train a US-specific (i.e., only trained with US English part of the corpus) student model in three settings: 1) trained from scratch with no guidance from the teacher (No-Teacher), 2) trained student model with $\lambda=0.5$, 3) trained student model with $\lambda=0.9$. We examine two settings of student models to investigate how much the supervision of the teacher influences the student alignments, however, a setting with $\lambda=1.0$ is not reported because this setting does not result in better ASR performance. In all experiments, default values of $\lambda$ and $T$ are  0.9 and 4, respectively (unless otherwise specified). 

\begin{table}[ht!]\caption{Percentage of CSO between the multi-accent model (Teacher model) and different US-specific models.}\label{tab:alignment}
\centering
\scalebox{0.85}{
\begin{tabular}{l p{.7cm}p{0.7cm}p{0.7cm}p{0.7cm}p{.7cm}p{0.7cm}}
\hline
& \multicolumn{2}{c}{\textbf{No\_Teacher}} & \multicolumn{2}{c}{\textbf{Teacher\_$\lambda$:0.5}} &  \multicolumn{2}{c}{\textbf{Teacher\_$\lambda$:0.9}} \\
\textbf{ } & \textbf{Train} & \textbf{Test} & \textbf{Train} & \textbf{Test}  & \textbf{Train} & \textbf{Test } \\ \hline \hline
\textbf{Teacher model} & 81\% & 78\% & 89\% & 82\% & 95\% & 87\%   \\ 
\hline

\end{tabular}
}
\end{table}

To examine the overlap between spikes of two CTC trained models, we obtain the index of maximum character per frame for each model, resulting in two sequences of character indexes for each utterance. Next, we calculate an average of overlap between these two sequences for all utterances of the data, referred to as the "characters' spikes overlap" (CSO). Table\ref{tab:alignment} shows the percentages of CSO for training and test utterances of US English data between the baseline and the three accent-specific models. The baseline and No\_Teacher model have approximately 80\% CSO, showing the difference between CTC trained models' spikes. However, training the student model under the guidance of the baseline model increases CSO to 95\% for the train data. The proposed approach could increase CSO to an acceptable point for knowledge distillation where we only need to have aligned models for training data. However, despite the CSO increase for the test data, there exists some room for improvement in scenarios of ensembling the CTC models in the evaluation step.

\begin{table}[ht!]\caption{CER\% of adapting a multi-accent trained model to Indian accent using tempered KL-divergence with three soft-outputs: Outputs of the multi-accent model (MA), outputs of accent-specific model with no teacher (No-Teacher) and an accent-specific student model with $\lambda$= 0.9 and T=4 (Student). The baseline performance is shown in the first row. }\label{tab:adaptation}
\centering
\scalebox{1}{
\begin{tabular}{l c c c}
\hline
& \multicolumn{3}{c}{\textbf{MA (baseline: 14.2) + Adaptation}}    \\ 
\hline
\textbf{ } & \textbf{MA \cite{adaptation}} & \textbf{No-Teacher} & \textbf{Student} \\ \hline \hline
\textbf{MA adapted} & 12.4 & 12.2 & 11.7   \\ 
\hline

\end{tabular}
}
\end{table}

To investigate whether having high CSO influences knowledge transferring between two models, we adapt the baseline model to Indian accent using tempered KL-divergence \cite{adaptation} with three different soft-outputs (Table \ref{tab:adaptation}). Adapting the baseline model using outputs of the model itself as soft-outputs  \cite{adaptation} results in a +12.7\% relative CER improvement compared to the baseline. However, adapting using the student model outperforms the former setting, demonstrating that accent-specific models provide better soft-outputs which represent the underlying similarities between characters of that accent.  Using outputs of No-Teacher model as soft outputs of the adaptation process performs better than using the baseline's soft outputs, demonstrating the efficacy of accent-specific models even with low CSO.

\subsection{Improved multi-accent model with accent-specific teachers}

To examine our proposed multi-step knowledge transferring (Figure \ref{fig:framework}), first, we train a multi-accent model with all accents pooled together without any teacher information (MA\_NT). To this end, we train a student model with the same architecture to examine how much the single teacher model can regularize another multi-accent model (Figure \ref{fig:framework}-b). Having fixed soft character alignments from the teacher model not only generalizes the student model (MA\_ST), it makes the training more stable compared to CTC which changes the alignments dynamically during the training steps. As shown in Table \ref{tab:proposed}, the resulting MA\_ST model achieves improved performance compared to MA\_NT model.  To consider the proposed two steps of knowledge transferring in the single teacher case, we exploit  MA\_ST to teach an even more improved student model (MA\_ST1). This MA\_ST1 achieves a greater decrease in CER for all accents, demonstrating the efficacy of knowledge distillation over multiple teacher-student generations.

\begin{table}[!th]
\caption{\label{tab:proposed} {\it CER\% of multi-accent without teacher model (MA\_NT), Multi-accent with a single teacher mode (MA\_ST), multi-accent with accent specific teacher (MA\_MT) and accent specific models on US English, Hispanic English (HIS) and Indian English (IND).    }}

\begin{center}
\begin{tabular}{ p{1.5cm} c c c c c}
    \hline
        Model & Teacher Model &  US  & HIS & IND  & Ave \\
      \hline
     MA\_NT &None&  14.1 & 13.5 &  14.2& 13.9 \\
     ACC\_SP0& None& 17.3& 18.2& 17.6& 17.7\\
	  \hline
	 ACC\_SP & MA\_NT &  15.2&  14.5 &  13.2 & 14.3\\
	 MA\_ST  & MA\_NT& 11.9 & 11.6 & 12.0 & 11.8\\
	 \hline
	 MA\_ST1  & MA\_ST& 11.5 & \textbf{10.8} & 11.7 & 11.3\\
	 MA\_MT & Acc\_Sp& 11.3 & \textbf{10.8}& 11.4 & 11.2\\
	 \hline
	 Acc\_Sp1& MA\_MT & 14.2& 14.0 & 12.7& 13.6\\
	 MA\_MT1 & Acc\_Sp1& \textbf{11.2} & \textbf{10.8}& \textbf{11.3}& \textbf{11.1}\\
	 \hline
	 MA\_MT1\_Adpt & MA\_MT1& 11.2 &10.8 &10.9 & 11\\
	 MA\_MT1\_Adpt1& Acc\_Sp1& \textbf{11.2}& \textbf{10.7}& \textbf{10.5}& \textbf{10.8} \\
	 \hline

\end{tabular}
\end{center}
\end{table}

In this section, we investigate the effectiveness of having multiple accent-specific teacher models to train a multi-accent model.  We train three accent specific models from scratch as well as under the guidance of MA\_NT. Each model is only trained with the corresponding accent speech data (ACC\_SPs models).  These resulting accent specific models perform better than accent specific models which are trained with no teacher (ACC\_SP vs. ACC\_SP0 in Table \ref{tab:proposed}). These improvements show that using soft-outputs of the teacher model not only aligns the student models (Table \ref{tab:alignment}), it regularizes them as well.

We leverage the three accent-specific aligned models (i.e., ACC\_SP\_US,  ACC\_SP\_HIS, and ACC\_SP\_IND) to train an overall improved multi-accent multi-teacher model (MA\_MT). As shown in Figure \ref{fig:framework}, soft-outputs of each accent data are produced by the corresponding ACC\_SP model. The resulting model outperforms both former improved multi-accent models (MA\_ST and MA\_ST1), demonstrating that accent specific models provide better underlying relations between output characters, resulting in a more generalized student model. Comparing the average CER of ACC\_SP models (14.3\%)  with MA\_ST (11.8\%) demonstrates that for our multi-accent data, accent-specific teachers are more effective than a superior multi-accent teacher model. Following the diagram of Figure \ref{fig:framework}  to transfer knowledge one step further, we achieved the best multi-accent model (MA\_MT1) which outperforms all other multi-accent models yielding a relative CER gain of +20.1\% vs. baseline (MA\_NT). Adapting the best multi-accent model to each accent using both Acc\_Sp1 (see Figure \ref{fig:framework}) and MA\_MT1 to regularize the adaptation process, again supports the idea that accent specific model outputs perform better than multi-accent model outputs (MA\_MT1\_Adpt vs. MA\_MT1\_Adpt in Table \ref{tab:proposed}).  

\section{Conclusions}
In this study, we investigate employing student-teacher learning to advance an LSTM-CTC muli-accent model. We proposed to train multiple CTC accent-specific models under the guidance of a single multi-accent teacher model to align their outputs. This approach not only aligned the CTC trained model, but also generalized the resulting student models. To achieve an advanced multi-accent model, we also proposed a novel approach of knowledge transfer, where we trained a multi-accent model (student model) under the supervision of accent-specific aligned models. The proposed method was shown to significantly outperform the baseline multi-accent model trained without any teacher model. Having this advanced multi-accent model leads to further improvement, by training new accent-specific models from which we guide a new multi-accent model. This second step of knowledge transfer yields the best multi-accent model providing a +20.1\% CER gain over the original baseline multi-accent model. Accent-specific models not only led to the best multi-accent model, their soft-outputs regularize the adapting process of the multi-accent model to each accent. Finally, constraining the adapted model to imitate the accent-specific models' outputs results in a more generalized adapted model compared to the method employing outputs of the multi-accent model itself.

\bibliographystyle{IEEEbib}
\bibliography{strings,refs}

\end{document}